\documentclass[adp,fleqn]{w-art}
\usepackage{times}
\usepackage{w-thm}

\usepackage[]{graphicx}
\begin{document}

\DOIsuffix{theDOIsuffix}
\Volume{NN}
\Issue{N}
\Copyrightissue{nn}
\Month{MM}
\Year{200Y}

\pagespan{1}{}
\Receiveddate{dd Mmmm 200Y}
\Reviseddate{Dd Mmmmm 200Y}
\Accepteddate{dD Mmmmmm 200Y}
\Dateposted{DD Mmmmmmm 200Y}

\title[Dark energy from quantum fluctuations]{Dark energy from quantum fluctuations}

\author[B. Broda]{Bogus{\l}aw Broda\footnote{Corresponding author\quad
E-mail:~\textsf{bobroda@uni.lodz.pl},
            Phone: +48\,42\,6355673,
            Fax: +48\,42\,6785770}}
\address{Department of Theoretical Physics, University of {\L}\'od\'z, Pomorska 149/153, PL--90--236 {\L}\'od\'z, Poland}

\author[M. Szanecki]{Micha{\l} Szanecki\footnote{E-mail:~\textsf{michalszanecki@wp.pl}}}

\begin{abstract}
We have derived the quantum vacuum pressure $p_{\rm vac}$ as a
primary entity, removing a trivial and a gauge terms from the
cosmological constant-like part (the zeroth term) of the effective
action for a matter field. The quantum vacuum energy density
$\tilde{\varrho}_{\rm vac}$ appears a secondary entity, but both
are of expected order. Moreover $p_{\rm vac}$ and
$\tilde{\varrho}_{\rm vac}$ are dynamical, and therefore they can
be used in the Einstein equations. In particular, they could
dynamically support the holographic dark energy model as well as
the ``thermodynamic'' one.
\end{abstract}
\maketitle                   







\noindent In \cite{Broda1,Broda2} we have proposed an estimation
method yielding a phenomenologically reasonable value of the
quantum vacuum energy density \cite{Weinberg,Carrol}. Not only the
very order of this value is in accordance with our expectations
but the very estimation procedure as well. More precisely, we
follow an old idea to use quantum vacuum fluctuations as a
physical constituent
of dark energy (cosmological constant) \cite{Zeldovich}.\\

\noindent Almost everybody knows the absurd, textbook estimation
of the quantum vacuum energy density \cite{Weinberg}, cited as an
example of a commonly acknowledged error,
\begin{align}
  \varrho_{\rm vac}\approx\frac{1}{4 \pi^{2}}\int\limits_{0}^{\Lambda_{\rm
  \textsc{uv}}}k^{3}\mathrm{d}k=\frac{1}{16 \pi^{2}}{\Lambda_{\rm
  \textsc{uv}}}^{4}\approx 3.4\times10^{94}\rm\,kg/m^3,
  \label{Wrong density 1}
\end{align}
where the UV cutoff $\Lambda_{\rm
  \textsc{uv}}=\Lambda_{\rm
  \textsc{P}}$ (the planckian momentum), and $c=\hbar=1$ for simplicity.
  Evidently, there is something wrong with the value \eqref{Wrong density
  1}, but what? One could, for example, argue that quantum vacuum energy
  does not couple to gravity but it would contradict the
  well-known arguments that all kinds of energy couple to gravity.
  Besides, the result $\varrho_{\rm vac}=0$ is not satisfactory
  either.\\

  \noindent An inspiring visual hint is implicitly given by Polchinski in
  Figure (6.)1 of his  paper \cite{Polchinski}. Following his
  hint, which we hopefully have done, we have isolated and removed a purely
  vacuum loop term giving rise to the absurd, huge value \eqref{Wrong density
  1}. Such a standpoint and a realization of the
  corresponding estimation has been already successfully
  elaborated in \cite{Broda1,Broda2}. Therefore, here we will
  only remind key steps of that procedure. The estimation
  method is limited only to the flat
  Friedmann--Lema\^{\i}tre--Robertson--Walker (FLRW) background geometry but it is
  sufficient for qualitative cosmological considerations.\\

  \noindent The full quantum cosmological constant-like contribution from a
  single bosonic mode is of the form \cite{Broda1,Broda2}
\begin{align}
  S_{\rm eff}=-\frac{1}{4}\frac{1}{(4\pi G)^{2}}\int
  \sqrt{-g}\;\mathrm{d}^{4}x,
  \label{Effective action 1}
\end{align}
 which is obviously of the order (one fourth, for minkowskian $g_{\mu\nu}$) of \eqref{Wrong density
 1}. Eq.~\eqref{Effective action 1} can be derived, e.g., from the
 Schwinger--DeWitt \cite{DeWitt} formula for the effective action of a free
 single bosonic field in an external gravitational background.
 Assuming the FLRW metric with the expansion parameter function $a(t)$, and
 next power expanding it around $a(0)=\nolinebreak 1$ we get
\begin{align}
  a(t)=1+\dot{a}(0)t+\frac{1}{2}\ddot{a}(0)t^{2}+\ldots\;,
  \label{a(t) Power expansion 1}
\end{align}
 and consequently
\begin{align}
  \sqrt{-g}=\sqrt{\left(1+2H_{0}t+(1-q_{0}){H_{0}}^2t^2+\ldots\right)^{3}}.
  \label{Sqrt(-g) Power expansion 1}
\end{align}
 The first term in the parentheses, number 1, corresponds to the
 trivial, purely vacuum, disconnected loop, and it can be removed. We
 should strongly stress that it is not an ad hoc step but a
 standard procedure in quantum field theory. It appears that
 by virtue of a gauge transformation generated by
\begin{align}
  \xi_{\mu}=\left(\frac{1}{2}H_{0}{\mathbf{x}}^2,
-H_{0}tx^{i}\right),
  \label{Gauge transformation generating function 1}
\end{align}
 the term linear in $t$ can also be removed.
 Here we have assumed standard definitions for the Hubble expansion rate
 $H_{0}$ and the deceleration parameter $q_{0}$:
\begin{align}
 H_{0}\equiv\frac{\dot{a}(0)}{a(0)},\;\;\;q_{0}\equiv
 -{H_{0}}^{-2}\ddot{a}(0).
  \label{Hubble and deceleration parameters definitions}
\end{align}
 Then,
\begin{align}
 \int L_{\rm eff}\;\mathrm{d}t\approx -\frac{1}{4}\frac{1}{(4\pi
 G)^{2}}\int
 \frac{3}{2}\left(1-q_{0}\right){H_{0}}^{2}t^{2}\;\mathrm{d}t.
  \label{Effective Lagrangian time integration 1}
\end{align}
Since our considerations are, by construction, limited to an
infinitesimal time $t$ we can write
\begin{align}
 L_{\rm eff}\approx -\frac{1}{4}\frac{1}{(4\pi
 G)^{2}}\lim_{T\rightarrow T_{\rm
 P}}\frac{1}{T}\int\limits_{0}^{T}
 \frac{3}{2}\left(1-q_{0}\right){H_{0}}^{2}t^{2}\;\mathrm{d}t,
  \label{Effective Lagrangian time averaging 1}
\end{align}
where we have physically interpreted the infinitesimal time as the
planckian time $T_{\rm P}$. Finally,
\begin{align}
 L_{\rm eff}\approx -\frac{1}{4}\frac{1}{(4\pi
 G)^{2}}\frac{1}{2}\left(1-q_{0}\right){H_{0}}^{2}{T_{\rm
 P}}^{2}=-\frac{1}{128\pi^{2}G}\left(1-q_{0}\right){H_{0}}^{2}.
  \label{Effective Lagrangian time averaging 2}
\end{align}
Numerically, Eq.~\eqref{Effective Lagrangian time averaging 2}
yields
\begin{align}
 \left|L_{\rm eff}\right|\sim 0.01\; \varrho_{\rm
 crit}/\mbox{mode},
  \label{Effective Lagrangian absolute value 1}
\end{align}
a very realistic result.\\

\noindent A new important observation we would like to present in
this article is the possibility to rewrite Eq.~\eqref{Effective
Lagrangian time averaging 2} without the subscript ``0'' (at $H$
and $q$) which bounds our considerations to our present time
instant. We are entitled to do so because in our calculations
there is nowhere explicit reference to our present epoch. In other
words, we can perform the time expansion around $t=0$ at any time
instant because our present time instant is by no means
distinguished in our estimation. This observation is really
important because it means that Eq.~\eqref{Effective Lagrangian
time averaging 2} can be used not only to recover current value of
the quantum vacuum energy density but to describe its dynamics as
well.\\

\noindent First of all, we should notice that, strictly speaking,
\eqref{Effective Lagrangian time averaging 2} is the quantum
vacuum pressure rather the quantum vacuum energy density, as could
seem at first sight. It simply follows exactly from the same
(analogous) argumentation which says that energy density is equal
to (minus) lagrangian density for time-independent fields. This
time the fields are spatially-independent ($\partial_{i}\phi=0$),
and therefore emerges the pressure instead of the energy density.
Namely,
\begin{align}
 p_{i}\equiv
 T_{ii}=\frac{\partial\mathcal{L}}{\partial\partial^{i}\phi}\partial_{i}\phi-g_{ii}\mathcal{L}=-g_{ii}\mathcal{L}\approx\mathcal{L}.
  \label{Pressure and Lagrange density 1}
\end{align}
Summarizing all above, we have
\begin{align}
 p_{\rm
 vac}=-\frac{N}{128\pi^{2}G}\left(1-q\right)H^{2}=-\frac{N}{128\pi^{2}G}\left[\frac{\ddot{a}}{a}+\left(\frac{\dot{a}}{a}\right)^{2}\right],
  \label{Vacuum pressure 1}
\end{align}
where $N$ is an effective number of modes. E.g.\ $N=N_{\rm
B}-N_{\rm F}$, i.e.\ $N$ is the difference between the number of
bosonic and
fermionic fundamental modes (see, also \cite{Broda3} and \cite{Broda4}).\\

\noindent Using the Einstein (acceleration) equation
\begin{align}
 -qH^{2}\equiv\frac{\ddot{a}}{a}=-\frac{4\pi G}{3}(\varrho+3p),
  \label{Einstein equation usage 1}
\end{align}
we can derive the quantum vacuum energy density
$\tilde{\varrho}_{\rm vac}$ from the quantum vacuum pressure
$p_{\rm vac}$. Namely
\begin{align}
 \tilde{\varrho}_{\rm
 vac}=\frac{3}{8\pi
 G}\left[\frac{N}{16\pi}+q\left(2-\frac{N}{16\pi}\right)\right]H^{2}.
  \label{Vacuum density with einstein EQ 1}
\end{align}
Here the barotropic coefficient $\tilde{w}_{\rm vac}$ is not
constant but $q$-dependent, i.e.\
\begin{align}
 \tilde{w}_{\rm vac}\equiv\frac{p_{\rm vac}}{\tilde{\varrho}_{\rm vac}}=-\frac{1}{3}\cdot\frac{1}{1+\frac{32\pi
 q}{N(1-q)}}.
  \label{Barotropic coefficient 1}
\end{align}
For example, for $\tilde{w}\sim -1$ and $q\sim -1$ we get
$N\sim 100$, quite a realistic result.\\

\noindent One could wonder if such a simple use of
Eq.~\eqref{Einstein equation usage 1} actually reproduces valid
$\tilde{\varrho}_{\rm vac}$ from $p_{\rm vac}$. In a limited
sense, i.e.\ when there are no other sources of gravitational
field, it is really so. One needs the pressure $p_{\rm vac}$ or
the energy density $\tilde{\varrho}_{\rm vac}$ to insert it to one
of the Einstein equations. One can work with $p$ or with
$\tilde{\varrho}$, but for consistency, to be sure that classical
einsteinian calculations will be identical, Eq.~\eqref{Einstein
equation usage 1} should be satisfied. Therefore, in spite of the
fact that only true $p_{\rm vac}$ is directly accessible one can
consistently, though in an above limited context, use
$\tilde{\varrho}_{\rm vac}$ coming from Eq.~\eqref{Einstein
equation usage 1} as an equivalent of a true
quantum vacuum energy density. This limited equivalency is denoted by ``$\sim$'', but we can still work with a generally valid $p_{\rm vac}$ if necessary. \\

\noindent Rewriting Eq.~\eqref{Vacuum density with einstein EQ 1}
in terms of $H$ and $\dot{H}$ we obtain
\begin{align}
 \tilde{\varrho}_{\rm
 vac}=\frac{3}{8\pi
 G}\left[\left(\frac{N}{8\pi}-2\right)H^{2}+\left(\frac{N}{16\pi}-2\right)\dot{H}\right],
  \label{Vacuum density with einstein EQ 2}
\end{align}
which is exactly the form of the vacuum energy density postulated
in the framework of (an extended) holographic dark energy model in
\cite{Granda}. The holographic expression
\begin{align}
 \varrho_{\rm
 hol}=\frac{3}{8\pi
 G}\left(\alpha H^{2}+\beta \dot{H}\right),
  \label{Vacuum density holographic 1}
\end{align}
contains two, in principle, arbitrary parameters $\alpha$ and
$\beta$. The origin of \eqref{Vacuum density holographic 1} in the
framework of the holographic principle is in a sense
``kinematical'', i.e.\ it lacks any dynamical support. Therefore
our analysis could serve as a dynamical explanation of
Eq.~\eqref{Vacuum density holographic 1} with $\alpha$ and $\beta$
fixed by dynamics, i.e.\
\begin{align}
\alpha=\frac{N}{8\pi}-2\;\;\mbox{and}\;\;\beta=\frac{N}{16\pi}-2.
  \label{Dynamcial holographic coefficients 1}
\end{align}
One could also use our dynamical approach as a basis for Volovik's
thermodynamic, qualitative considerations \cite{Volovik}. Namely,
using some general thermodynamic arguments and analogies coming
from condense matter physics Volovik argues that measured quantum
vacuum energy density should be almost zero. Moreover, according
to him, it should constantly run to zero in the course of the
evolution of the Universe. Obviously, these qualitative
considerations are not capable to yield any quantitative result.
We like his point of view, and we think that, in a sense, our
infinitesimal expansion around any consecutive time instant could
be interpreted as realization of his idea.\\

\noindent We have shown that our primary approach
\cite{Broda1,Broda2} aimed to estimate the current value of the
quantum vacuum energy density $\varrho_{\rm vac}^{0}$ can be
successfully extended to the dynamical expression \eqref{Vacuum
density with einstein EQ 1}. Moreover, we have indicated a
possibility to use our approach as a dynamical basis of the
holographic and Volovik's ``thermodynamic'' approaches.

\begin{acknowledgement}
This work has been supported by the University of {\L}\'od\'z.
\end{acknowledgement}


\begin{thebibliography}{11}

\bibitem{Broda1}
Broda, B., Bronowski, P., Ostrowski, M.\ \& Szanecki, M.,
{\it``Vacuum Driven Accelerated Expansion''}, Ann.\ Phys.\
(Berlin), \textbf{17}, 855--863, 2008; [arXiv: 0708.0530].

\bibitem{Broda2}
Broda, B.\ \& Szanecki, M., {\it``Quantum Vacuum and Accelerated
Expansion''}, {\it``Dark Energy and Dark Matter: Observations,
Experiments and Theories''}, EAS Publications Series \textbf{36},
167--171 (2009); [arXiv: 0812.4892].

\bibitem{Weinberg}
Weinberg, S., {\it The Cosmological Constant Problem}, Rev.\ Mod.\
Phys.\ \textbf{61}, 1--23 (1989).

\bibitem{Carrol}
Carroll, S.\ M., {\it The Cosmological Constant}, Liv.\ Rev.\
Rel.\ \textbf{4} (2001); [arXiv: astro-ph/0004075].

Padmanabhan, T., {\it Cosmological Constant---the Weight of the
Vacuum}, Phys.\ Rep.\ \textbf{380}, 235--320 (2003); [arXiv:
hep-th/0212290].

Padmanabhan, T., {\it Dark Energy: Mystery of the Millennium}, AIP
Conf.\ Proc.\ \textbf{861}, 179--196 (November 3, 2006) Albert
Einstein Century International Conference; [arXiv:
astro-ph/0603114].

\bibitem{Zeldovich}
Zel`dovich, Y.\ B., {\it Cosmological Constant and Elementary
Particles}, JETP Lett.\ \textbf{6}, 316--317 (1967), translated
from Zh.\ Eksp.\ Teor.\ Fiz., Pis`ma Redaktsiyu, \textbf{6}
883--884 (1967).

\bibitem{Polchinski}
Polchinski, J., {\it``Rapporteur Talk: The Cosmological Constant
and the String Landscape''}, {\it``THE QUANTUM STRUCTURE OF SPACE
AND TIME}, Proc.\ of the 23rd Solvay Conference on Physics
Brussels, Belgium, 1--3 December 2005; 216--236 (World Scientific,
2007); [arXiv: hep-th/0603249].

\bibitem{DeWitt}
DeWitt, B.\ S., {\it Quantum Field Theory in Curved Spacetime},
Phys.\ Rep.\ \textbf{19}, 295--357 (1975).

DeWitt, B.\ S., {\it The Global Approach to Quantum Field Theory},
(Clarendon Press, 2003).

\bibitem{Broda3}
Broda, B.\ \& Szanecki, M., \textit{``Induced Gravity and Gauge
Interactions Revisited''}, Phys.\ Lett.\ B \textbf{674}, 64--68
(2009); [arXiv: 0809.4203].

\bibitem{Broda4}
Broda, B.\ \& Szanecki, M., {\it``Vacuum Pressure, Dark Energy and
Dark Matter''}, [arXiv: 0906.5078].

\bibitem{Granda}
Granda, L.\ N., Cardona, W.\ \& Oliveros, A., {\it``Current
Observational Constraints on Holographic Dark Energy Model''},
[arXiv: 0910.0778].

Granda, L.\ N.\ \& Oliveros A., {\it ``Infrared Cut-off Proposal
for the Holographic Density''}, Phys.\ Lett.\ B \textbf{669},
275--277 (2008); [arXiv: 0810.3149].

Granda, L.\ N.\ \& Oliveros A., {\it ``New Infrared Cut-off for
the Holographic Scalar Fields Models of Dark Energy''}, Phys.\
Lett.\ B \textbf{671}, 199--202 (2009); [arXiv: 0810.3663].

\bibitem{Volovik}
Volovik, G.\ E., {\it Vacuum Energy: Myths and Reality}, Int.\ J.\
Mod.\ Phys.\ D \textbf{15}, 1987--2010 (2006); [arXiv:
gr-qc/0604062].

Volovik, G.\ E., {\it Cosmological Constant and Vacuum Energy},
Ann.\ Phys.\ (Leipzig) \textbf{14}, 165--176 (2005); [arXiv:
gr-qc/0405012].


\end{thebibliography}
\end{document}